\documentclass[twocolumn,showpacs,preprintnumbers,amsmath,amssymb, prl]{revtex4-1}

\usepackage{stmaryrd}
\usepackage{txfonts}
\usepackage{amssymb}
\usepackage{mathrsfs}
\usepackage{graphicx}
\usepackage{dcolumn}
\usepackage{bm}
\usepackage{epsfig}
\usepackage{color}                    
\usepackage{hyperref}                 

\begin{document}
\preprint{\href{http://dx.doi.org/10.1103/PhysRevB.86.180506}{S.-Z. Lin, L. N. Bulaevskii, and C. D. Batista , Phys. Rev. {\bf{B}} {\bf 86}, 180506(R) (2012).}}

\title{Vortex Dynamics in Ferromagnetic Superconductors: Vortex Clusters, Domain Walls and Enhanced Viscosity}

\author{Shi-Zeng Lin, Lev N. Bulaevskii and Cristian D. Batista}

\affiliation{Theoretical Division, Los Alamos National Laboratory, Los Alamos, New Mexico 87545, USA}

\begin{abstract}
We demonstrate that there is a long-range vortex-vortex attraction in ferromagnetic superconductors due to polarization of the magnetic moments. Vortex clusters are then stabilized in the ground state for low vortex densities.  The motion of vortex clusters driven by the Lorentz force excites magnons. This regime becomes unstable at a threshold velocity above which domain walls are generated for slow relaxation of the magnetic moments and the vortex configuration becomes modulated. This  dynamics of vortices and magnetic moments can be probed by transport measurements. 
\end{abstract}


\date{\today}

 \maketitle

\noindent {\it Introduction --} Superconductivity (SC) and magnetism are at the heart of modern condensed matter physics. While they seem to be antagonist according to the standard BCS theory,  a large family of magnetic superconductors were discovered in the last decades. 
Examples include coexistence of antiferromagnetism or helical ferromagnetic (FM) order in ternary superconducting compounds\cite{Bulaevskii85}, uniform ferromagnetism in triplet superconductors\cite{Saxena00,Aoki01,Pfleiderer01}, and antiferromagnetism in the Re$\rm{Ni_2B_2C}$ borocarbides\cite{Canfield98} (Re represents a rare earth element) and in the recently discovered iron-based superconductors\cite{Chu09}. The interplay between SC and magnetism allows to control the superconducting properties through the magnetic subsystem, and vice versa. 
These phenomena open new possibilities for  applications to superconducting electronics and magnetic storage devices\cite{Buzdin04,Lyuksyutov05}.

The Abrikosov vortices of superconductors are a natural link between the superconducting condensate and the magnetic moments. 
Vortices are induced either by external magnetic fields or by the MMs \cite{Tachiki80}. 
On the other hand, the magnetic subsystem supports collective spin-waves and topological excitations that are domain walls. 
Because vortices are magnetic objects, they are expected to interact strongly with MMs via Zeeman coupling. Indeed, as we discuss below,  vortex motion can drive magnetic domain walls.

The MMs provide a novel handle to control the vortex behavior in the static and dynamic regimes. It was demonstrated that magnetic domains induce a vortex pinning that is 100 
times stronger than the one induced by columnar defects\cite{Bulaevskii00}. In the flux flow regime,  vortex motion radiates magnons by  transferring energy  into the magnetic system. This effect has been
recently proposed by Shekhter \emph{et al.} for antiferromagnetic superconductors  \cite{Shekhter11}. By assuming a rigid vortex lattice and fast relaxation of the MMs, it is demonstrated that 
Cherenkov radiation of magnons occurs when the vortex lattice velocity, $\mathbf{v}$, satisfies $\mathbf{G}\cdot \mathbf{v}=\Omega(\mathbf{G})$, where $\mathbf{G}$ is the vortex lattice wave vector and $\Omega(\mathbf{G})$ is the magnon dispersion. This emission gives an additional contribution to the vortex viscosity that manifests as a voltage drop  in the I-V characteristics. Thus,  the overall 
dissipation is reduced for a given current. Vortex motion can also be used to probe the spectrum of excitations in the magnetic subsystem.\cite{Bulaevskii05}

Several questions remain to be addressed. It is known that intrinsic nonlinear effects  of the magnetic subsystem become important for high energy magnon excitations.  However, it is unclear if magnon excitations remain stable in this nonlinear regime. On the other hand, the interaction between the magnetic subsystem and vortices may become comparable or even stronger 
than the inter-vortex repulsion. Therefore, the vortex lattice may be modified by this effect. Finally, the dominant dissipation mechanism of vortices when domain walls are excited by the vortex motion is unknown.

Here we study the vortex dynamics in FM superconductors. The Zeeman coupling between vortices and MMs induces an additional vortex-vortex attraction  that is comparable to the inter-vortex repulsion for a large enough magnetic susceptibility. This attraction leads to the formation of vortex clusters at low vortex densities. We also show that magnetic domain walls are created when vortex clusters driven by the Lorenz force reach a threshold velocity. The interaction between domain walls and vortices greatly enhances the vortex viscosity and causes hysteresis in the dynamics of the whole system. The vortex configuration is modulated by the domain walls.

\vspace{2mm}
\noindent {\it Model--} Uniform FM order and SC suppress each  other because of the exchange and electromagnetic coupling between the MMs and Cooper pairs\cite{Bulaevskii85}. However they
could coexist  in triplet FM superconductors \cite{Saxena00,Aoki01,Pfleiderer01}, such as $\rm{UGe_2}$,  layered magnetic superconductors consisting FM and SC layers \cite{Sumarlin92,McLaughlin99}, such as $\rm{Sm_{1.85}Ce_{0.15}CuO_4}$, or artificial bilayer systems\cite{Lyuksyutov05,Buzdin05}. Here we study the vortex dynamics in these FM superconductors. An applied dc magnetic field  perpendicular to the ferromagnetic easy  axis creates a vortex lattice that is driven by a dc in-plane current. We use the approximation of straight vortex lines and the description of vortices is two dimensional. 

The total Gibbs free energy functional of the system, in terms of the vector potential $\mathbf{A}$, magnetization $\mathbf{M}$ and vortex position $\mathbf{R_i}=(x_i, y_i)$, is
\begin{equation}\label{eqb1} 
{G}(\mathbf{A}, \mathbf{M}, \mathbf{R}_i)=d\int d r^2\left({g}_{\text{sc}}+{g}_M+{g}_{\text{int}}\right)+\frac{1}{8\pi}\int_{\text{out}} d r^3\mathbf{B}^2, 
\end{equation}
where $d$ is the thickness of the system and the last term is the magnetic energy outside the superconductor. The energy functional density for the SC subsystem in the London approximation is
\begin{equation}\label{eqb2}
{g}_{\text{sc}}(\mathbf{A})=\frac{\mathbf{B}^2}{8\pi }-\frac{\mathbf{B}\cdot \mathbf{H}_{\text{ext}}}{4\pi }+\frac{1}{8\pi\lambda _L^2}\left(\frac{\Phi_0}{2\pi}\nabla\phi-\mathbf{A} \right)^2,
\end{equation}
with $\mathbf{B}=\nabla\times \mathbf{A}$. $\phi$ is the superconducting phase, $H_{\rm{ext}}$ is the applied magnetic field, $\lambda_L$ is the London penetration depth and $\Phi_0=h c/2e$ is the flux quantum. The energy functional density of the magnetic subsystem is 
\begin{equation}\label{eqb3}
{g}_M= \frac{J}{2}(\nabla  \mathbf{M})^2-\frac{J_A}{2}M_x^2,
\end{equation}
where $J$ and $J_A$ are the exchange and anisotropy parameters. The easy axis is taken along the $x$ direction. We assume that the magnitude of the magnetic moment is conserved, $|M|=M_s$, 
where $M_s$ is the saturated magnetization value. Because of the anisotropy, the magnetic Hamiltonian has two degenerate minima and supports stable domain walls. The Zeeman interaction between MMs and SC is
\begin{equation}\label{eqb4}
g_{\text{int}}=- \mathbf{B}\cdot\mathbf{M}.
\end{equation}
The vortex axis is taken along the $z$ direction. The straight vortex lines approximation is valid when $d\ll \lambda_L$ or $d\gg \lambda_L$.  The spreading of magnetic field associated with vortices near the surface of superconductors has to be taken into account for $d \sim \lambda_L$, \cite{Kirtley99}. By minimizing $g_{\rm{sc}}+g_{\rm{int}}$ with respect to $\mathbf{A}$, we obtain the magnetic field associated with vortices
\begin{equation}\label{eqb5}
\lambda _L^2\nabla \times \nabla \times (\mathbf{B}-4\pi  \mathbf{M})+\mathbf{B}=\Phi _0\sum_i\delta \left(\mathbf{r}-\mathbf{R}_i\right)\hat{\mathbf{z}}.
\end{equation}
$M_z(k)=B_z(k)\tilde{\chi}_{zz}(k)$ in the linear response region when $M_z/M_s\ll 1$. As $\lambda_L$ is much larger the magnetic correlation length $\xi_m\sim \sqrt{J/J_A}$, we can use a local approximation for  $\tilde{\chi}_{zz}(k) \simeq 1/ J_A=\chi_0/(1+4\pi \chi_0)$. The uniform susceptibility $\chi_0 \propto \left\langle M_z({\boldsymbol k=0})M_z({\boldsymbol k=0})\right\rangle$ diverges at $J_A=4\pi$, which signals an instability of the magnetic  subsystem.  The FM ordering along the $x$-direction coexists with superconductivity only when $J_A>4\pi$. \cite{Blount1979}

According to Eq.\eqref{eqb5}, the magnetic field of a vortex at $\mathbf{R}_i$ is 
\begin{equation}\label{eqb6}
B_z\left(\mathbf{k}, \mathbf{R}_i\right)=\frac{\Phi _0}{1+\lambda _e^2 \mathbf{k}^2}\exp(i \mathbf{k}\cdot \mathbf{R}_i),
\end{equation}
with a renormalized penetration depth $\lambda _e\equiv\lambda _L/\sqrt{1+4\pi  \chi _{0}}$.

\begin{figure}[t]
\psfig{figure=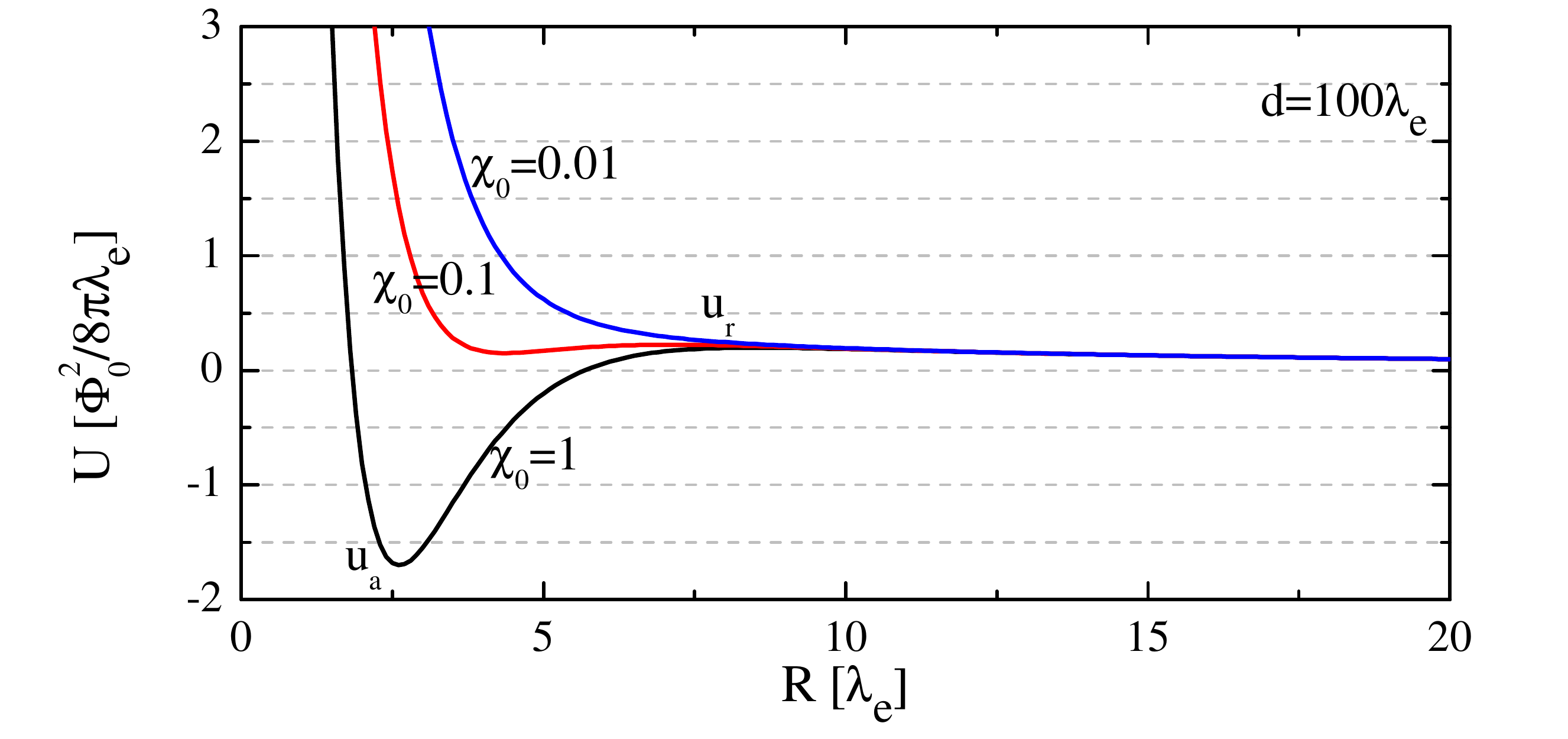,width=\columnwidth}
\caption{\label{f1}(color online) Vortex-vortex interaction potential for different values of $\chi_0$ according to Eqs. (\ref{eqb7}) and (\ref{eqb9}). Attraction is induced due to the Zeeman coupling between vortices and MMs, and the long-range repulsion arises from the electromagnetic fields outside the SC.}
\end{figure}

\vspace{2mm}
\noindent {\it Attraction between vortices via MMs --} We calculate now the interaction between two vortices at $\mathbf{R}_i$ and $\mathbf{R}_j$. Vortices interact with each other through the exchange of massive photons described by ${g}_{\rm{sc}}$, which leads to a short-range repulsion. As was first discussed by Pearl, vortices also interact through the exchange of massless photons outside the SC, as described by the last term in Eq. (\ref{eqb1}). This contribution leads to a long-range repulsion\cite{Pearl64,Wei96}. The total repulsion energy is
\begin{equation}\label{eqb7}
U_r(R)=\frac{\Phi _0^2d}{8\pi ^2 \lambda_e ^2}K_0\left(\frac{{R}}{\lambda_e }\right)+\frac{\Phi _0^2}{8\pi  \Lambda}\left[H_0\left(\frac{R}{\Lambda }\right)-Y_0\left(\frac{R}{\Lambda }\right)\right],
\end{equation}
with $\mathbf{R}\equiv\mathbf{R}_i-\mathbf{R}_j$ and $\Lambda=2\lambda_e  \coth [d/\lambda_e]$ is the modified Pearl length. $K_i$ is the modified Bessel function, $H_0$ is the Struve function and $Y_0$ is the Weber function.

\begin{figure*}[t]
\psfig{figure=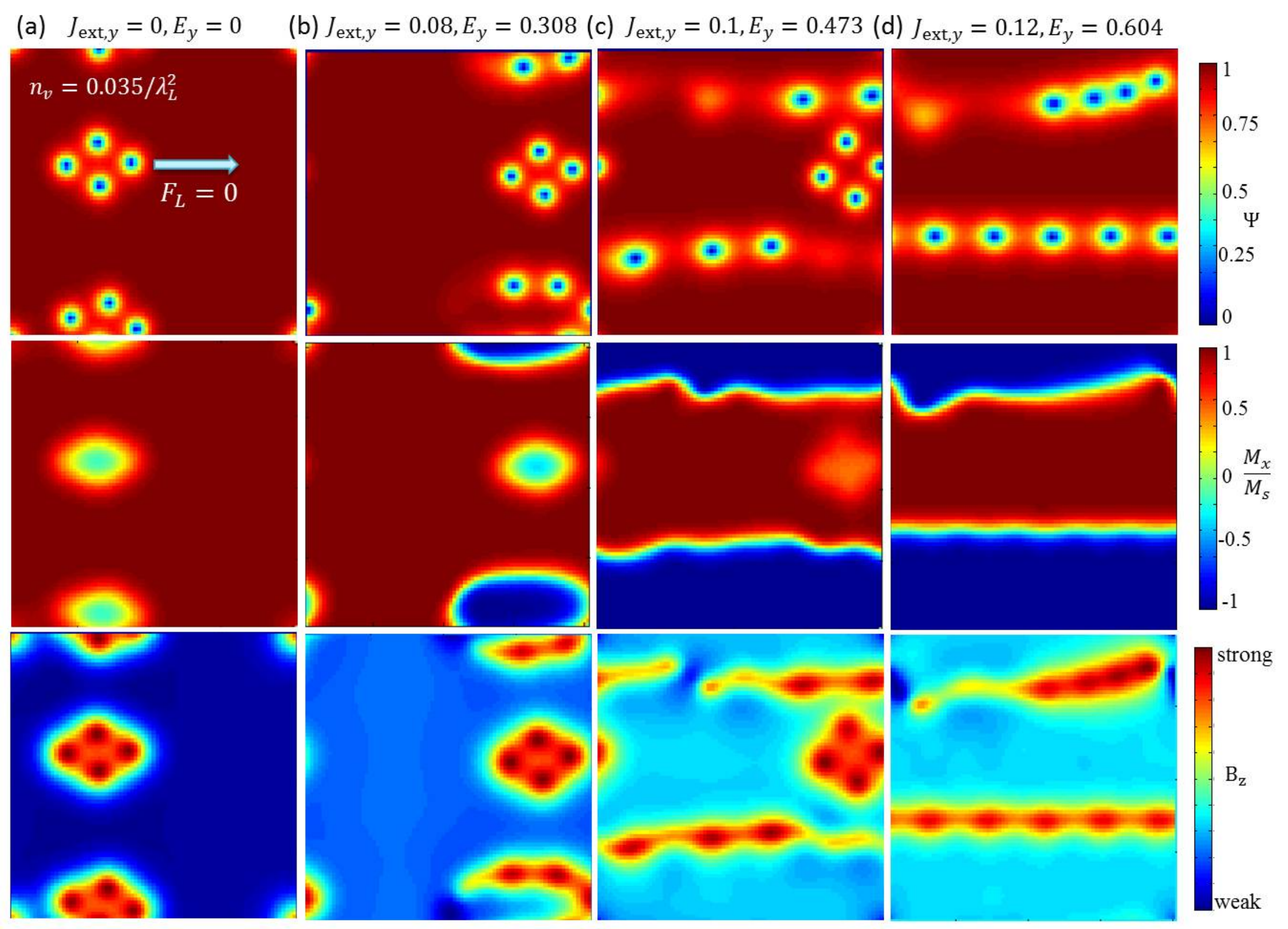,width=17.0cm}
\caption{\label{f2}(color online) (a-b): Development of the amplitude of superconducting order parameter $|\Psi|$, magnetic structure ($x$-component of the magnetic moment: $M_x$) and $B_z$ as the current increases. The vortex positions  correspond to the regions with suppressed $|\Psi|$. (a) Static configuration  with $J_{\rm{ext},y}=0$. In (b-d), domain walls are created and the vortex configuration is modulated.  $|\Psi|$ is suppressed (top row) and $B_z$ is maximal (bottom row) in the normal core of vortices. MMs are canted by vortices so $M_x$ is reduced (middle row).}
\end{figure*}

A vortex at $\mathbf{R}_i$ polarizes the surrounding MMs. This effect leads to an effective attraction to a vortex at $\mathbf{R}_j$. The magnetic energy due to the presence of vortices is $d\int dr^2 (g_M+g_{\text{int}})$ with $B_{z}^v=B_z(R_i)+B_z(R_j)$ and $M_{z}^v=\chi_0 B_{z}^v/(1+4\pi\chi_0)$. The contribution from the gradient term in Eq. (\ref{eqb3}) is much smaller than the anisotropic contribution because $k \xi_m\ll 1$ with $k\sim 1/\lambda_e$. By using $M_x^2+M_z^2=M_s^2$, we obtain the attractive interaction
\begin{equation}\label{eqb9}
U_a(R)=-\frac{d}{2} \int d r^2B_{z}^vM_{z}^v=-\frac{d\chi _0\Phi _0^2 R   }{4\pi(1+4\pi\chi_0)  \lambda_e ^3}K_1\left(\frac{R}{\lambda_e }\right).
\end{equation}
In the presence of attraction, the repulsion through the electromagnetic fields outside the SC in Eq. (\ref{eqb7}) cannot be neglected because it prevents the formation of a single cluster. The physics here is similar to the laminar phase in conventional type I superconductors\cite{TinkhamBook}.

The effect of finite velocity, $\mathbf{v}$, on the vortex-vortex interaction is negligible because $\tilde{\chi}_{zz}$ depends weakly on $\mathbf{v}$ for $\xi_m/\lambda_e\ll 1$.  The attractive component is comparable to the repulsion for $\chi_{0}\sim 1$ and the energy minimum takes place at ${R}_m\sim\lambda_e$. Fig. (\ref{f1}) shows the energy of two vortices separated by a distance ${R}$. For $\chi_0\sim 1$, the net interaction is attractive for large separations $\lambda_e<R< \Lambda$ and repulsive at short distances $R<\lambda_e$. There is also a long-range repulsion for $R>\Lambda$ due to the surface effect. Since the susceptibility $\chi_0$ decreases with $J_A$, the attractive component drops as anisotropy increases. 
The inter-vortex  interaction becomes purely repulsive for $\chi_0\ll 1$.

\vspace{2mm}
\noindent {\it Excitation of domain walls --} We  introduce the equation of motion for MMs and vortices that is used in the numerical simulation. The FM subsystem is described by the Landau-Lifshitz-Gilbert equation\cite{Gilbert04}
\begin{equation}\label{eq1}
\partial _t \mathbf{m}=-\gamma \mathbf{m}\times \mathbf{B}_{\text{eff}}+\alpha  \mathbf{m}\times \partial _t\mathbf{m},
\end{equation}
where $\gamma$ is the gyromagnetic ratio, $\mathbf{m}=\mathbf{M}/M_s$ is the normalized MM, $\alpha$ is the damping coefficient and the effective magnetic field is $\mathbf{B}_{\text{eff}}=-\delta [g_M+g_{\rm{int}}]/\delta \mathbf{M}$. The vortex subsystem is described by the time-dependent Ginzburg-Landau equations
\begin{equation}\label{eqTDGL1}
\frac{\hbar ^2}{2m D}\partial _t\Psi =-\left[\alpha_s\Psi+\beta \left|\Psi\right|^2\Psi+\frac{\hbar^2}{2m}\left({i }\nabla +\frac{2\pi}{\Phi_0}\mathbf{A}\right)^2\Psi\right],
\end{equation}
\begin{equation}\label{eqTDGL2}
\frac{\sigma}{c}\partial _t\mathbf{A} = \mathbf{J}_s+\mathbf{J}_{\rm{ext}}-\frac{c}{4\pi}\nabla\times(\nabla\times\mathbf{A}-4\pi\mathbf{M}),
\end{equation}
with the supercurrent
\begin{equation}\label{eqTDGL4}
\mathbf{J}_s=\frac{{{e}\hbar }}{{i{m}}}({\Psi ^*}\nabla \Psi  - \Psi \nabla {\Psi ^*}) - \frac{{4e^2}}{{{m}c}}|\Psi {|^2}{\mathbf{A}},
\end{equation}
$D$ is the diffusion coefficient, $\sigma$ is the  conductivity in the normal state, $\mathbf{J}_{\rm{ext}}$ is the external current and other parameters are defined according to the usual convention. The MMs stop responding to the vortex motion when the average magnetic field, $\bar{B}_z\approx n_v \Phi_0$ with $n_v$ being the vortex density, is larger than the saturation value, $B_s=M_s J_A$, and the two subsystems become decoupled. Therefore, we shall consider the interesting region $\bar{B}_z<B_s$.

In the long wavelength and weak damping $\alpha\ll1$ limits, the magnon dispersion for the FM system of Eq.\eqref{eq1} is
\begin{equation}\label{eqvs}
\Omega^2=\omega_0^2+v_s^2 k^2, \ \ \ v_s=\gamma M_s\sqrt{\left(2-m_{\text{z0}}^2\right)J_A J},
\end{equation}
\begin{equation}\label{eqomega}
\omega_0^2=J_A^2\gamma ^2 M_s^2\left(1-m_{\text{z0}}^2\right)\left[1 +i { \alpha }{(2-m_{\text{z0}}^2)}{( 1-m_{\text{z0}}^2 )^{-1/2}}\right],
\end{equation}
where $m_{\text{z0}}$ is the $z$ component of the MMs in the ground state and $v_s$ is the magnon velocity. $\text{Re}(\omega_0)$ is the energy gap and $\text{Im}(\omega_0)$ is the magnon relaxation rate.  $\text{Re}(\omega_0)=100\rm{\ GHz}$ and $v_s=50\rm{\ m/s}$ for typical ferromagnets.\cite{Pickart67}

We then establish general relations of the energy transfer between MMs and vortices. The vortex velocity acquires an ac part, ${\tilde {\mathbf v} }_i$, because of the interaction between vortices and MMs, $\mathbf{v}_i=\bar{\mathbf{v}}+\tilde{\mathbf{v}_i}$. The energy balance for the whole system reads
\begin{equation}\label{eq5}
\eta \bar{v}^2+ \eta\left\langle\tilde{v_i}^2\right\rangle_{i,t} +\frac{1}{n_v} \frac{\alpha}{M_s\gamma}\left\langle \int dr^2 (\partial _t\mathbf{M})^2\right\rangle_{x,t}=\mathbf{F}_L \cdot\bar{\mathbf{v}},
\end{equation}
where $\left\langle \cdots\right\rangle_{i,t}$ denotes average over vortices and time, and $\left\langle \cdots\right\rangle_{x,t}$ denotes average over space and time. The first and second term on the left-hand side (lhs) correspond to Bardeen-Stephen (BS) damping with coefficient $\eta=\Phi_0^2\sigma/(2\pi c^2\xi^2)$ \cite{TinkhamBook}, where $\xi=\sqrt{{\hbar ^2}/{(2m|\alpha_s|)}}$ is the coherence length. The third term on the lhs accounts for the dissipation due to precession of MMs. The term on the right-hand side is the work done by  the Lorentz force $F_L$. The effective viscosity $\eta_{\rm{eff}}=F_L/\bar{v}$ is enhanced due to the interaction between vortices and MMs,
\begin{equation}\label{eq5aa}
\eta_{\rm{eff}}=\eta+\frac{\eta}{\bar{v}^2}\left\langle\tilde{v_i}^2\right\rangle_{i,t} +\frac{1}{n_v \bar{v}^2}\frac{\alpha}{M_s\gamma}\left\langle \int dr^2 (\partial _t\mathbf{M})^2\right\rangle_{x,t}.
\end{equation}
 Off resonance, the contribution of the magnetic damping is small, thus $\bar{\mathbf{v}}\approx \mathbf{F}_L/\eta$. Since $F_L=J_{\rm{ext}}\Phi_0/c$ and $E=\bar{v} n_v \Phi_0/c$ with an external current $J_{\rm{ext}}$ and electric field $E$, the underlying dynamics can be probed by the I-V measurement.

The effect of magnons on the vortex dynamics depends on the vortex density. When the average inter-vortex distance is smaller than the value corresponding to the potential minimum, $n_v<1/R_m^2$, the attraction between vortices dominates.  Vortices form circular clusters with internal triangular structure in the ground state, as shown in Fig. \ref{f2} (a) obtained from our simulations\cite{simulation}. The distance between neighboring vortices inside the cluster is of order $\lambda_e$, and the separation between neighboring clusters is of order $\sqrt{\pi R_c^2/(n_v\lambda_e^2)}$, with a cluster radius given by $R_c\approx\Lambda[-u_a/(3u_r)]^{1/3}$.\cite{supplement} The attractive, $u_a<0$, and repulsive, $u_r>0$, energies are defined in Fig.~\ref{f1}. The vortex clusters start to merge and more complex vortex configurations, such as stripes, are possible for larger values of $n_v$. The $H=H_{c1}$ transition from the uniform Meissner state to the state with vortex clusters is of first order \cite{Tachiki79,Buzdin91,szlin11a} in contrast to the second order phase transition expected for conventional type II superconductors\cite{TinkhamBook}. Vortex clusters in conventional superconductors with inter-vortex attraction, such as Nb, have been observed experimentally, see Ref. \cite{Brandt95} for a review.  

For finite transport current, each cluster driven by the Lorentz force moves as a whole and  polarizes the MMs along its way. The MMs relax to their positions of equilibrium after the vortex cluster leaves that region. The polarization and excitation of magnons, and subsequent relaxation of MMs thus causes vortex dissipation through the magnetic subsystem\cite{Bulaevskii12a}. The static structure of the vortex clusters remains the same for a small $v$  because the change of the vortex-vortex interaction is negligible for $\xi_m/\lambda_e\ll 1$.

Here we derive a resonant condition between the motion of vortex clusters and magnon emission. The magnetic field distribution produced by the vortex motion has a dominant wave vector $G_x=2\pi/R_m$, with $R_m\approx \lambda_e$ as shown in Fig. \ref{f1}. The unperturbed ordered state has $M_{z0}=0$. The resonant condition $G_x v=\Omega(G_x)$ gives a resonant velocity for vortices moving along the $x$ direction
\begin{equation}\label{eq9}
v_t=\gamma M_s\sqrt{2J_A J+\frac{R_m^2 J_A^2}{4\pi^2}}.
\end{equation}
This linear analysis is correct as long as the canted MMs satisfy the condition that $M_{zc}\approx\Phi_0/(J_A R_m^2)\ll M_s$ [or $J_A\gg \Phi_0/(R_m^2M_s)$].

\begin{figure}[t]
\psfig{figure=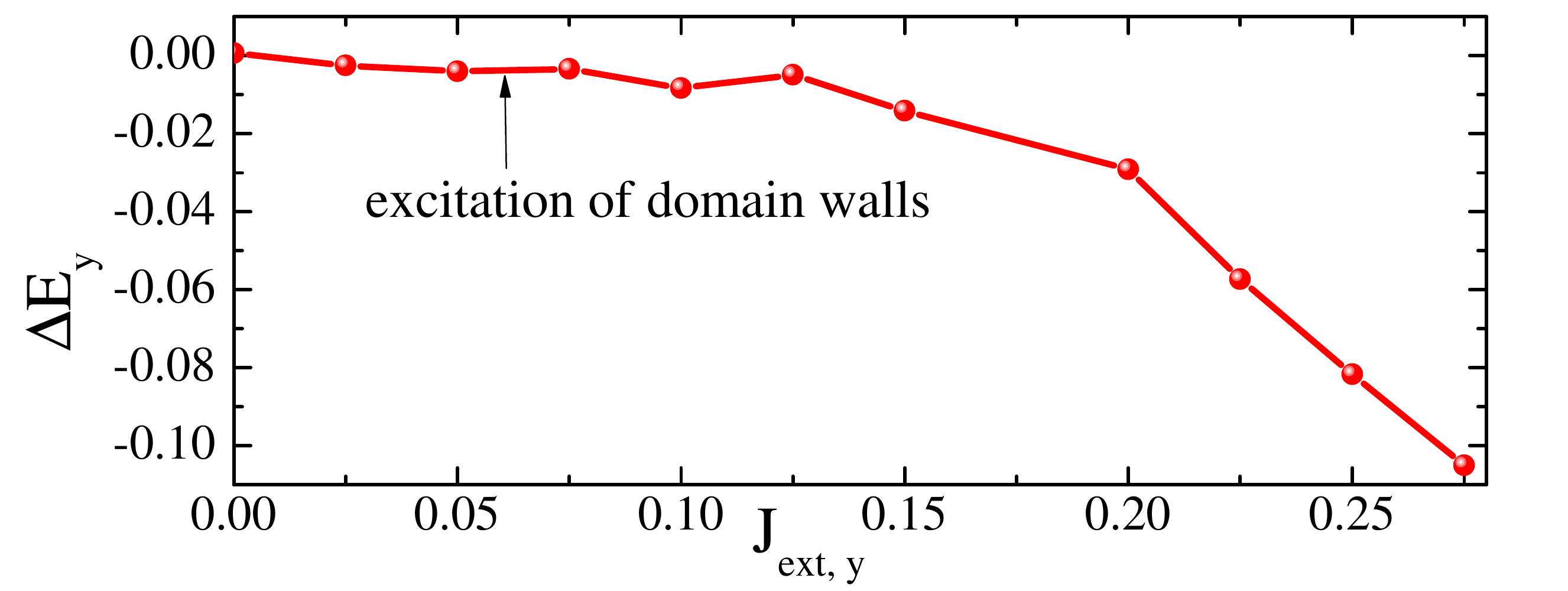,width=\columnwidth}
\caption{\label{f3}(color online) Difference between the electric fields induced with and without magnetic moments as a function of current $J_{\rm{ext},y}$, $\Delta E=E_M-E_B$, where $E_M$ is the electric field for the system with magnetic moments and $E_B$ is the electric field for the system without magnetic moments. The vortex viscosity increases when domain walls are created resulting in a drop of the electric field (vortex velocity).}
\end{figure}

The oscillation amplitude of MMs and the ac part of the  vortex velocity are greatly enhanced in resonance and $\eta_{\rm{eff}}$ increases  according to Eq. (\ref{eq5aa}). Two competing processes are involved in the magnetic subsystem: the energy input from vortex motion and the magnetic relaxation. For large dissipation ($\alpha\gg 1$), the excited magnon is quickly dissipated and the vortex cluster with canted MMs remains stable. On the contrary, the incoming energy accumulates for  weak magnetic dissipation, $\alpha\ll 1$, and increases with time. This effect leads to an instability of the magnon excitations that has been discussed decades ago both experimentally \cite{Bloembergen72} and theoretically\cite{Suhl57,Schlomann61,ChenBook}. For a large enough oscillation amplitude, the MMs are no longer restricted to one of the symmetry-breaking states (there are two degenerate ground states with $M_{x0}=\pm M_s\sqrt{1-m_{z0}^2}$) and they can flip to the other ground state (with opposite $M_{x0}$). Domain walls are then created  as shown in Fig. \ref{f2}(b, c, d). $M_z$ becomes large inside the domains walls and this effect increases the coupling between the magnetic subsystem and vortices.  For $v\gg v_t$, the cluster structure evolves into vortex stripes along the driving direction [\ref{f2}(b, c, d)]. The domain walls are oriented along the vortex stripes due to the strong attraction between vortices and  domain walls. Vortex stripes for large driving forces and random pinning potentials have also been observed in numerical simulations without MMs \cite{Reichhardt03}. As vortex clusters drive domain walls, the dissipation increases and the vortex velocity (voltage) drops as shown in Fig.~\ref{f3}. The threshold velocity obtained from simulations where the domain walls are created is compatible with that estimated from Eq.(\ref{eq9}).

\vspace{2mm}
\noindent {\it Discussions --} The magnetic susceptibility is small, $\chi_{0}\ll1$, in bulk FM superconductors such as $\rm{UGe_2}$\cite{Huxley03}. Thus, the attraction between vortices is negligible and the ground state is a triangular vortex lattice. In the flux flow regime, the vortex lattice is resonant with the oscillations of MMs when $\mathbf{G}\cdot \mathbf{v}=\Omega(\mathbf{G})$ is satisfied. We predict an enhancement of the vortex viscosity at resonance, which can be probed by the I-V measurement. A large susceptibility, $\chi_{0}\sim 1$, is needed to realize the vortex cluster configuration. This requirement can he fulfilled by some cuprate superconductors with rear-earth elements (Re), such as $\rm{ReBa_2Cu_3O_x}$, where $\rm{Re}$ ions order antiferromagnetically below $T_N \sim 1\ \rm{K}$.  Spins are free from the molecular field above the Neel temperature $T_N\sim 1\ K$ and can be easily polarized \cite{Hor1987,Allenspach1995} to mediate the attraction between vortices in the low magnetic field region.  The vortex cluster phase can also be achieved in heterostructures of superconductors and ferromagnets with large susceptibility\cite{Yang04}. On the other hand, random pinning centers  may prevent the formation of vortex clusters   because pinning is strong for a small vortex densities. However,  
vortex motion in the flux flow regime quickly averages out the effect of random pinning centers\cite{Koshelev94,Besseling03} and the cluster structure may be recovered.

\vspace{2mm}

 \noindent {\it Acknowledgement --} We are indebted to V. Kogan, B. Maiorov, M. Weigand, C. J. Olson Reichhardt and C. Reichhardt for helpful discussions. The present work is supported by the Los Alamos Laboratory directed research and development program with project number 20110138ER.


\begin{thebibliography}{41}
\bibitem [{sim()}]{simulation}%
  \BibitemOpen
  \href@noop {} {}\bibinfo {note} {For numerical purposes,length is in units of
  $\lambda_L$, time $t$ is in units of $\xi^2/D$, conductivity $\sigma$ is in
  units of $c^2\tau/(4\pi\lambda_L^2)$, the superconducting order parameter
  $\Psi$ is in units of $\sqrt{|\alpha_s|/\beta}$, the magnetic field
  $\mathbf{B}$ and magnetization $\mathbf{M}$ are in units of
  $\Phi_0/(2\pi\xi^2\kappa)$, $\gamma$ is in units of $2\pi\kappa D/\Phi_0$ and
  the exchange parameter $J$ is in units of $\lambda_L^2$. Here
  $\lambda_L=\sqrt{{m c^2\beta}/{(16\pi |\alpha_s| e^2)}}$ is the London
  penetration depth, $\xi=\sqrt{{\hbar ^2}/{(2m|\alpha_s|)}}$ is the coherence
  length, $\kappa=\lambda_L/\xi$ is the Ginzburg-Landau parameter. The system
  is discretized into a mesh with size $0.2\lambda_L$. We use periodic boundary
  conditions\cite{szlin11a}. Equation (\ref{eq1}) is solved by an explicit
  numerical scheme developed in Ref. \cite{Serpico01}, and Eqs.
  ~(\ref{eqTDGL1}) and (\ref{eqTDGL2}) are solved by using the
  finite-difference method in Ref.~\cite{szlin11a}. We apply a current along
  the $y$ direction, $J_{\rm{ext},y}$ and calculate the electric field along
  the same direction $E_y=-\partial_t A_y$. The I-V curve is calculated with
  and without the magnetic subsystem to obtain the magnetic contribution to the
  voltage drop. The simulation parameters are: anisotropy parameter $J_A=13$,
  saturation magnetization $M_s=0.05$, exchange parameter $J=4$, $\alpha=0.01$,
  $\gamma=0.5$, $\sigma=1$ and $\kappa=2$. The size of simulation box is
  $16\lambda_L\times 16\lambda_L\times 1\lambda_L$ and density of vortex is
  $n_v=0.035/\lambda_L^2$.}\BibitemShut {Stop}%
\end{thebibliography}
%

\newpage
\appendix
\section{Supplement : Characterization of the vortex cluster phase}
Here we characterize the vortex cluster phase by using a simple model. The vortex clusters form a triangular lattice with lattice constant
$a_c$. Each cluster contains $n_c$ vortices. In the dilute vortex phase, the interaction between vortex clusters is Coulomb-like, 
\begin{equation}\label{apeq1}
U_{\text{cc}}=\frac{\Phi _0^2n_c^2}{4\pi ^2 }\sum _{i,j}\frac{1}{R_{c,i}-R_{c,j}},
\end{equation}
where $R_{c,i}$ is the position of the cluster and the summation is over all clusters. The summation is calculated numerically and the result is well approximated by the expression
\begin{equation}\label{apeq2}
U_{\text{cc}}\approx \frac{\Phi _0^2n_c^2}{4\pi ^2 a_c}\left(1.6N_c^{1.5}-2N_c\right),
\end{equation}
where $N_c$ is the number of clusters. $N_c=n_vL^2/n_c$ and $a_c=L/\sqrt{N_c}$ in a sample with lateral size $L^2$ and vortex density $n_v$. The repulsion between all clusters is
\begin{equation}\label{apeq3}
U_{\text{cc}}=-\frac{\Phi _0^2n_v\sqrt{n_v}L^2}{2\pi ^{3/2}}\frac{R_c}{\lambda _s}+1.6\frac{\Phi _0^2\left(n_vL^2\right)^2}{4\pi ^2
L },
\end{equation}
where we have used $N_c=n_vL^2/\left(\pi  R_c^2/\lambda _s^2\right)$. $R_c$ is the cluster radius and $\lambda _s$ is the separation between
two nearest vortices inside the cluster. The number of vortices in a cluster is $n_c=\pi  R_c^2/\lambda _s^2$. Since the $R_c$-independent term of Eq.(\ref{apeq3}) is irrelevant in the following calculations, we will neglect it. 

\begin{figure}[t]
\psfig{figure=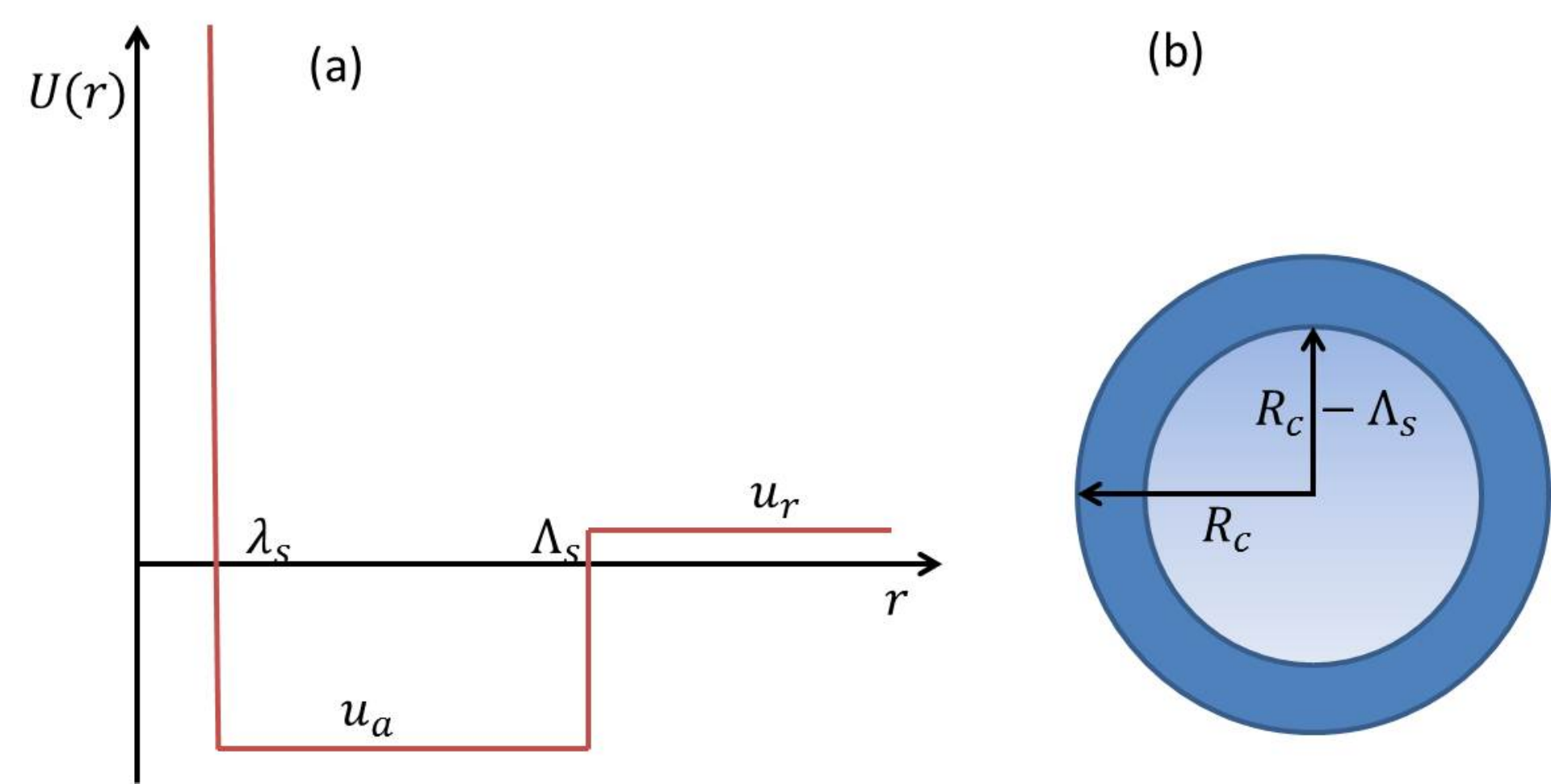,width=\columnwidth}
\caption{\label{fa1}(color online) (a) Approximated inter-vortex potential inside a vortex cluster. $\lambda _s\approx \lambda _e$ and $\Lambda _s=\Lambda$ compared
with the interaction profile in Fig.~1 of the main text. (b) Partition of the vortex cluster with radius $R_c$ to calculate the energy of the cluster.}
\end{figure}

To estimate the interaction energy inside the cluster we approximate the inter-vortex interaction as shown in Fig. \ref{fa1}(a). A vortex in the region with radius $R_c-\Lambda _s$ attracts vortices in a region $\pi \Lambda _s^2$ and repels vortices in a region $\pi  \left(R_c^2-\Lambda _s^2\right)$  [see Fig. \ref{fa1}(b)]. Then, the interaction energy in the region with radius $R_c-\Lambda _s$ is
\begin{equation}\label{apeq4}
U_{\text{c1}}=\frac{\pi }{2 \lambda _s^4} \left(R_c-\Lambda _s\right){}^2\left[\pi \Lambda _s^2u_a+\pi \left(R_c^2-\Lambda _s^2\right)u_r\right],
\end{equation}
where $u_a<0$ is the attraction and $u_r>0$ is the repulsion. A vortex in the ring $R_c-\Lambda _s<r<R_c$ attracts less vortices. The attraction region for a vortex in the ring can be written as $\alpha'  \Lambda _s^2$, with $\alpha' \approx 2$ obtained by direct integration over the ring area. The repulsion region is given by $\pi  R_c^2-\alpha'  \Lambda _s^2$. The interaction energy in the ring is then given by
\begin{equation}\label{apeq5}
U_{\text{c2}}=\frac{\pi  R_c}{\lambda _s^4} \Lambda _s\left[2 \Lambda _s^2u_a+\left(\pi  R_c^2-2 \Lambda _s^2\right)u_r\right].
\end{equation}
The total interaction in the vortex clusters is
\begin{equation}\label{apeq6}
U_c=N_c\left(U_{\text{c1}}+U_{\text{c2}}\right).
\end{equation}

In thick superconductors with $d>>\lambda$, we have $\left|u_a\right|>>u_r$ and $R_c>>\Lambda _s$. Then the energy of the whole system (apart from the $R_c$-independent contribution) is
\begin{equation}\label{apeq7}
U\approx -\frac{\Phi _0^2n_v\sqrt{n_v}L^2}{2\pi ^{3/2}}\frac{R_c}{\lambda _s}+\left(\frac{2\pi -\pi ^2}{\lambda _s^4}R_c \Lambda _s^3u_a+\frac{\pi
^2}{2\lambda _s^4}R_c^4u_r\right)\frac{n_vL^2\lambda _s^2}{\pi  R_c^2},
\end{equation}
where the first term accounts for the long-range repulsion between vortex clusters, and the second term accounts for the interaction inside clusters.
The interaction between vortex clusters scales with the density as $n_v^{3/2}$ and the interaction inside the cluster scales as $n_v$. The first term can be neglected in the
dilute vortex case $n_v\lambda _s^2<<1$. By minimizing $U$ with respect to $R_c$, we obtain the radius of the vortex cluster
\begin{equation}\label{apeq8}
R_c\approx \left(\frac{\pi -2}{\pi }\frac{\left|u_a\right|}{u_r}\right)^{1/3}\Lambda _s,
\end{equation}
and the distance between nearest vortex clusters is
\begin{equation}\label{apeq9}
a_c=\sqrt{\frac{\pi }{n_v}}\frac{R_c}{\lambda _s}.
\end{equation}
By comparing the potential in Fig.~\ref{fa1}(a) to the potential in Fig.~1 of the main text, we know that $\lambda _s\approx \lambda _e$ and $\Lambda _s\approx\Lambda$. Thus, we arrive to the results shown in the main text.

\end{document}